\documentclass{PoS}
\usepackage{wrapfig}

\title{
\vspace{-3cm}
\hspace{7cm}{\normalsize UTCCS-P-36, HUPD-0705, KANAZAWA-07-16} \\
\vspace{2cm}
Application of chiral perturbation theory to 2+1 flavor lattice QCD
 with $O(a)$-improved Wilson quarks}

\ShortTitle{Application of chiral perturbation theory}

\author{PACS-CS Collaboration: 
 \speaker{D. Kadoh}${}^{a}$\thanks{E-mail: kadoh@ccs.tsukuba.ac.jp}\ ,
 S.~Aoki${}^{b,c}$,
 N. Ishii${}^{a}$,
 K.-I.~Ishikawa${}^{d}$,
 N.~Ishizuka${}^{a,b}$,
 T. Izubuchi${}^{c,e}$,
 K.~Kanaya${}^{b}$,
Y. Kuramashi${}^{a,b}$,
Y. Namekawa${}^{a}$,
 M.~Okawa${}^{d}$,
 K. Sasaki${}^{a}$,
 Y.~Taniguchi${}^{a,b}$,
 A.~Ukawa${}^{a,b}$,
 N. Ukita${}^{a}$,
 T.~Yoshi\'e${}^{a,b}$
 \\
 \llap{${}^a$}Center for Computational Sciences, University of Tsukuba, Tsukuba, Ibaraki 305-8577, Japan\\
 \llap{${}^b$}Graduate School of Pure and Applied Sciences, University of Tsukuba, Tsukuba, Ibaraki 305-8571, Japan\\
 \llap{${}^c$}Riken BNL Research Center, Brookhaven National Laboratory, Upton, New York 11973, USA\\
 \llap{${}^d$}Department of Physics, Hiroshima University, Higashi-Hiroshima, Hiroshima 739-8526, Japan\\
 \llap{${}^e$}Institute for Theoretical Physics, Kanazawa University, Kanazawa, Ishikawa 920-1192, Japan}
 
\abstract{We apply chiral perturbation theory 
to the pseudoscalar meson mass and decay constant data 
obtained in the PACS-CS Project toward 
 2+1 flavor lattice QCD simulations with the $O(a)$-improved Wilson quarks. 
We examine the existence of chiral logarithms in the quark mass range 
from $m_{\rm ud}=47$MeV down to 6~MeV on a $(2.8 \ {\rm fm})^3$ box 
with the lattice spacing $a=0.09$fm.  
Several low energy constants are determined. 
We also discuss the magnitude of finite size effects based on 
chiral perturbation theory.
}

\FullConference{The XXV International Symposium on Lattice Field Theory\\
		 July 30-4 August 2007\\
		 Regensburg, Germany}

\begin{document}

\section{Introduction}
The PACS-CS Collaboration has been pushing $N_F=2+1$ full QCD simulations 
toward the physical point with the nonperturbatively 
$O(a)$-improved Wilson quark action and
the Iwasaki gauge action~\cite{ukawa,kuramashi,kuramashi07}.  
As presented in a separate report~\cite{ukita} in detail, 
sizable amount of hadron spectrum data has been collected at $\beta=1.9$ 
on a $32^3\times 64$ lattice down to the pion mass 
$m_\pi=210$MeV using the domain-decomposed
Hybrid Monte Carlo (DDHMC) algorithm~\cite{Luscher:2003vf}
 on the PACS-CS  computer. 
In this report we examine 
the chiral behavior of the pseudoscalar meson masses and decay constants 
in comparison with the prediction of chiral perturbation theory (ChPT). 
In particular, we focus on the three points: (i) signals for 
chiral logarithms, (ii) determination of low
energy constants in the chiral lagrangian, (iii) determination of 
the physical point with the ChPT fit.

\section{Simulation parameters}
We employ the $O(a)$-improved Wilson quark action with a
nonperturbative improvement coefficient $c_{\rm SW}=1.715$\cite{Aoki:2005et} and the
Iwasaki gauge action at $\beta=1.9$ on a $32^3\times 64$ lattice.
In Table~\ref{tab:param} we summarize the hopping
parameters $(\kappa_{\rm ud},\kappa_{\rm s})$ and the results for the
pion masses and the unrenormalized quark masses.  The latter is 
defined  through the axial vector
Ward-Takahashi identity (AWI)  by 
\begin{eqnarray}
am^{\rm AWI}_q =\lim_{t\rightarrow \infty} \frac{\langle\nabla_4
  A_4^{\rm imp}(t) P(0)\rangle}{2\langle P(t)P(0)\rangle},
\end{eqnarray}
where $P$ is the pseudoscalar density and $A_4^{\rm imp}$ is
the nonperturbatively $O(a)$-improved axial vector current\cite{Kaneko:2007wh}.

For later use we also define the renormalized quark mass and the
pseudoscalar meson decay constant in the continuum ${\overline{\rm MS}}$
scheme as follows:
\begin{eqnarray}
m^{\overline{\rm MS}}_q&=&\frac{Z_A \left(1+b_A
  \frac{m^{\rm AWI}}{u_0}\right)}
{Z_P \left(1+b_P \frac{m^{\rm AWI}}{u_0} \right)}m^{\rm AWI}_q,\\
f_{\rm PS}&=&2\kappa u_0 Z_A \left(1+b_A\frac{m^{\rm AWI}_q}{u_0}\right) \frac{C_A^s}{C_P^s}\sqrt{\frac{2C_P^l}{m_{PS}}}.
\end{eqnarray}
Here $C_{A,P}^s$ are the amplitudes extracted from the correlation functions
$\langle A_4^{\rm imp}(t) P(0)\rangle$ and $\langle P(t)P(0)\rangle$
with the exponentially smeared source and the local sink, while
$C_P^l$ is from $\langle P(t)P(0)\rangle$ with the local source and
the local sink. The renormalization factors
$Z_{A,P}$ and the improvement coefficients $b_{A,P}$ are evaluated
perturbatively up to 
one-loop level\cite{Aoki:1998ar,Taniguchi:1998pf}with the
tadpole improvement.

\begin{table}[h]
\hspace{1cm}
\begin{tabular}{ccccccc}\hline
$\kappa_{\rm ud}$ & $\kappa_{\rm s}$  & $am_{\pi}$  & $am_{\rm ud}^{\rm AWI}$ & 
$am_{\rm s}^{\rm AWI}$ & MD time  \\
\hline
\hline
0.13700  &  0.13640  &  0.32196(62)  &  0.02800(20) & 0.04295(30) &  2000 \\
0.13727  &  0.13640  &  0.26190(66)  &  0.01895(13) & 0.04061(18) &  2000 \\
0.13754  &  0.13640  &  0.18998(56)  &  0.01020(11) & 0.03876(18) &  2500 \\ 
0.13770  &  0.13640  &  0.13591(88)  &  0.00521(9)  & 0.03767(10) &  2000 \\
0.13781  &  0.13640  &  0.08989(291) &  0.00227(16) & 0.03716(20) &   350 \\
\hline
0.13754  &  0.13660  &  0.17934(78)  &  0.00908(7)  & 0.03257(17) &   900 \\
\hline
\end{tabular}
\caption{Hopping parameters for the up-down and the strange
quarks together with the results
of $am_\pi$, $am_{\rm ud}^{\rm AWI}$ and $am_{\rm s}^{\rm AWI}$.
MD time is the number of trajectories multiplied by the trajectory length.}
\label{tab:param}
\end{table}

\section{Results}

\subsection{Comparison with the previous CP-PACS/JLQCD results}
We first compare our results with the previous CP-PACS/JLQCD results,
both of which are obtained with the same quark and gauge actions at
$\beta=1.9$ but on different lattice sizes: $32^3\times
64$ for the former and $20^3\times 40$ for the latter.
For the hopping parameters only the combination $(\kappa_{\rm
  ud},\kappa_{\rm s})=(0.13700,0.13640)$ is in common, 
which is the heaviest case in the PACS-CS results 
whereas it is the lightest one in the
CP-PACS/JLQCD results.

\begin{figure}[h!]
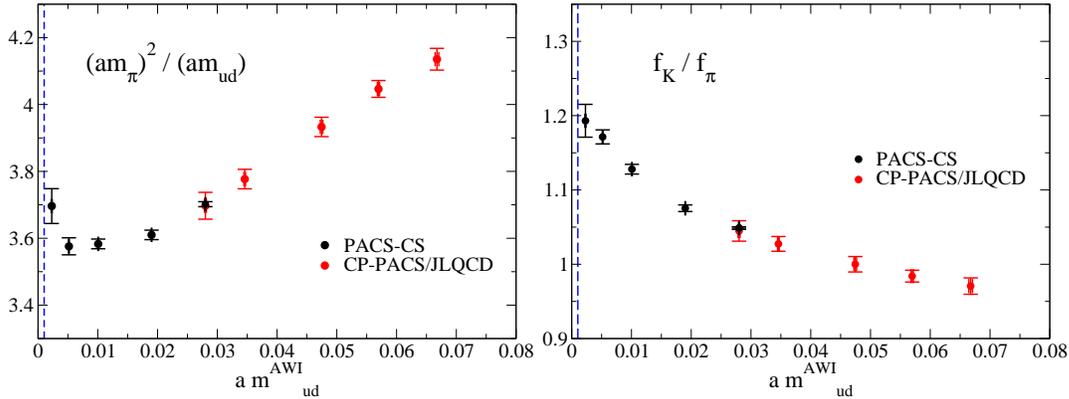

\begin{center}
\hspace{-0.6cm}
\includegraphics[width=7cm,keepaspectratio,clip]{f.mpi2bmud.cp.eps}
\includegraphics[width=7cm,keepaspectratio,clip]{f.fKfpi.cp.eps}
\caption{Comparison of the PACS-CS (black) and the CP-PACS/JLQCD (red)
  results for $(am_\pi)^2/(am_{\rm ud}^{\rm AWI})$ (left) and
  $f_K/f_\pi$ (right). $\kappa_{\rm s}$ is fixed at 0.13640.\vspace{-0.6mm}}
\label{fig:comparison}
\end{center}
\end{figure}

In Fig.~\ref{fig:comparison} we plot $(a m_\pi)^2/(am_{\rm ud}^{\rm AWI})$ 
and $f_K/f_\pi$ as a function of $am_{\rm ud}^{\rm AWI}$ with
$\kappa_{\rm s}$ fixed at 0.13640. The PACS-CS and the CP-PACS/JLQCD
results are denoted by the black and the red symbols, respectively.
The two sets of data together show a smooth behavior as 
a function
of $am_{\rm ud}^{\rm AWI}$, and at $\kappa_{\rm ud}=0.13700$  
($am_{\rm ud}^{\rm AWI}=0.028$) they show good consistency.

An important observation is that while the CP-PACS/JLQCD results show
an almost linear quark mass dependence 
both for $(a m_\pi)^2/(am_{\rm ud}^{\rm AWI})$ and $f_K/f_\pi$, 
we find a clear curvature for the PACS-CS results,
which is a characteristic 
feature of the ChPT prediction in the small quark mass region.
This curvature drives the PACS-CS results for $f_K/f_\pi$ close to the
experimental value toward the physical point.

\subsection{Chiral fit formulae}
In ChPT, the one-loop expressions for the pseudoscalar meson masses 
and the decay constants are given by\cite{gasser}
\begin{eqnarray}
m_{\pi}^2&=& 2 \hat m B_0 \left\{
1+\mu_\pi-\frac{1}{3}\mu_\eta 
+\frac{B_0}{f_0^2} \left(
16 \hat m  (2L_{8}-L_5) 
+16 (2 \hat m +m_{\rm s}) (2L_{6}-L_4)  
\right) \right\}, 
\label{eq:chpt_mpi}
\\
\hspace{-6mm}m_K^2&=&(\hat m +m_{\rm s})B_0 \left\{
1+\frac{2}{3}\mu_\eta
+\frac{B_0}{f_0^2}\left(
8(\hat m+m_{\rm s}) (2L_{8}-L_5)
+16(2 \hat m  +m_{\rm s}) (2L_{6}-L_4) 
 \right)\right\}, 
\label{eq:chpt_mk}
\\
f_\pi &=&f_0\left\{
1-2\mu_\pi-\mu_K
+ \frac{B_0}{f_0^2} \left (
8 \hat m L_5+8(2 \hat m +m_{\rm s})L_4
\right)\right\}, 
\label{eq:chpt_fpi}
\\
f_K &=&f_0\left\{
1-\frac{3}{4}\mu_\pi-\frac{3}{2}\mu_K-\frac{3}{4}\mu_\eta
+\frac{B_0}{f_0^2}\left(4(\hat m+m_{\rm s}) L_5+8(2 \hat m+ m_{\rm s})L_4 
\right)\right\}, 
\label{eq:chpt_fk}
\end{eqnarray}
where ${\hat m}=(m_{\rm u}+m_{\rm s})/2$ and 
$L_{4,5,6,8}$ are the low energy constants, and 
$\mu_{\rm PS}$ is the chiral logarithm defined by 
\begin{eqnarray}
\mu_{\rm PS}=\frac{1}{32\pi^2}\frac{m_{\rm PS}^2}{f_0^2}
\ln\left(\frac{m_{\rm PS}^2}{\mu^2}\right)
\label{eq:chlog}
\end{eqnarray}
with $\mu$ the renormalization scale.
There are six unknown low energy constants $B_0,f_0,L_{4,5,6,8}$ 
in the expressions above.
The low energy constants are scale-dependent so as to 
cancel that of the chiral logarithm (\ref{eq:chlog}). 
We determine these parameters                                  
by making a simultaneous fit for $m_\pi^2$, $m_K^2$, $f_\pi$ and $f_K$.



We also consider the contributions of the finite size effects based on ChPT.
At the one-loop level the finite size effect defined by 
$R_X=(X(L)-X(\infty))/X(\infty)$ for $X=m_\pi,m_K,f_\pi,f_K$ is given 
by~\cite{Colangelo:2005gd}:
\begin{eqnarray}
&& \ \ \ R_{m_\pi}
=\frac{1}{4}\xi_{\pi}\tilde g_1(\lambda_\pi)-\frac{1}{12}\xi_{\eta}\tilde g_1(\lambda_\eta),\\
&& \ \ \ R_{m_K} 
=\frac{1}{6}\xi_{\eta}\tilde g_1(\lambda_\eta),\\
&& \ \ \ R_{f_\pi}
=-\xi_{\pi}\tilde g_1(\lambda_\pi)-\frac{1}{2}\xi_{K}\tilde g_1(\lambda_K),\\
&& \ \ \ R_{f_K} 
=-\frac{3}{8}\xi_{\pi}\tilde g_1(\lambda_\pi)-\frac{3}{4}\xi_{K}\tilde g_1(\lambda_K)
-\frac{3}{8}\xi_{\eta}\tilde g_1(\lambda_\eta)
\end{eqnarray}
with
\begin{eqnarray}
\hspace{2.5cm} && \xi_{\rm PS} \equiv \frac{m_{\rm PS}^2}{(4\pi f_\pi)^2},
\quad \ \lambda_{\rm PS} \equiv m_{\rm PS} L, \quad
\tilde g_1(x)=\sum_{n=1}^{\infty}\frac{4m(n)}{{\sqrt n }x} K_1({\sqrt n} x), 
\end{eqnarray}
where $K_1$ is the Bessel function of the second kind and $m(n)$
denotes the multiplicities in the expression of $n=n_x^2+n_y^2+n_z^2$. 
With the use of these formulae
we estimate the possible finite size effects in our results.



\subsection{ChPT fits}
\begin{figure}[h!]
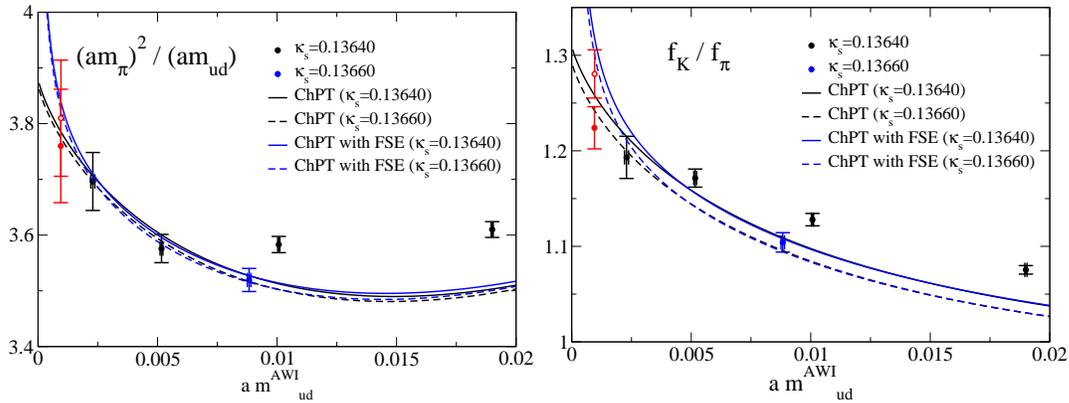

\begin{center}
\hspace{-0.6cm}
\includegraphics[width=7cm,keepaspectratio,clip]{f.mpi2bmud.eps}
\includegraphics[width=7cm,keepaspectratio,clip]{f.fKfpi.eps}
\caption{Fit results for $(am_\pi)^2/(am_{\rm ud}^{\rm AWI})$
  (left) and $f_K/f_\pi$ (right). Red solid (open) symbols denote the
 extrapolated values at the physical point by the ChPT formulae
without (with) the finite size effects. }
\label{fig:fit}
\end{center}
\end{figure}

We apply the ChPT formulae (\ref{eq:chpt_mpi})$-$(\ref{eq:chpt_fk})
to our results at four points 
$(\kappa_{\rm ud},\kappa_{\rm s})=(0.13781,$ $0.13640)$,
(0.13770,0.13640), (0.13754,0.13640), (0.13754,0.13660).  For these points, 
the $\rho$ meson mass satisfies the condition $m_\rho > 2m_\pi$.
The measured AWI qurak masses are used for ${\hat m}$ and $m_{\rm s}$ 
in eqs.(\ref{eq:chpt_mpi})$-$(\ref{eq:chpt_fk}).
The heaviest pion mass at $(\kappa_{\rm ud},\kappa_{\rm s})=(0.13754,0.13640)$
is about 430MeV with the use of the cutoff determined below.
The fit results are shown in Fig.~\ref{fig:fit}, 
where the black solid lines
are drawn with $\kappa_{\rm s}$ fixed at 0.13640 and the black dotted 
lines are for $\kappa_{\rm s}=0.13660$. 
The red solid symbols represent the extrapolated values at the
physical point whose determination is explained in Sec.~3.4. 
The heaviest point at 
$(\kappa_{\rm ud},\kappa_{\rm s})=(0.13754,0.13640)$
is not well described by ChPT both for 
$(a m_\pi)^2/(a m_{\rm ud}^{\rm  AWI})$ and $f_K/f_\pi$, and 
$\chi^2$/d.o.f. is rather large (see  Table~\ref{tab:fit}).

The results for the low energy constants are presented
in Table~\ref{tab:fit} where the
phenomenological values with the experimental
inputs\cite{Amoros:2001cp} and the MILC results\cite{Bernard:2006zp}
are given for comparison.
The renormalization scale is chosen to be $m_\eta=0.547$GeV.   
For $L_4$ and $L_5$ governing the behavior of $f_\pi, f_K$, 
our results show good agreement with both the
phenomenological estimates and the MILC results. 
On the other hand, some discrepancies are observed between three results 
for $2L_6-L_4$ and $2L_8-L_5$ which enter into the ChPT formulae 
for $m_\pi^2$ and $m_K^2$.

\begin{figure}[t!]
\vspace{3mm}
\begin{center}
\begin{tabular}{cc}
  \includegraphics[width=65mm]{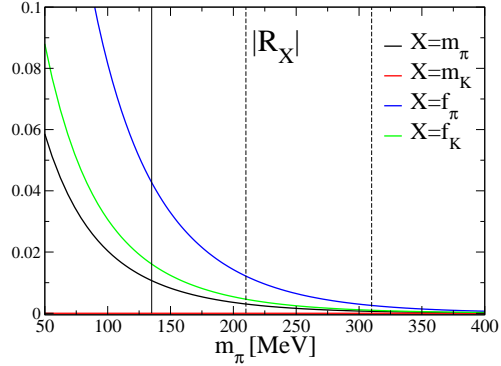}
\end{tabular}
\end{center}
\vspace{-.5cm}
\caption{$|R_X|$ for $X=m_\pi,m_K,f_\pi,f_K$ with $L=2.8$fm 
as a function of $m_\pi$.
Solid vertical line denotes the physical point and the dotted ones
are for our simulation points.}
\label{fig:ratio}
\end{figure}

In Fig.~\ref{fig:fit} we also draw the ChPT fit results 
incorporating the finite size effects. The blue solid
lines are drawn for $\kappa_{\rm s}=0.13640$ and the blue dotted ones
for $\kappa_{\rm s}=0.13660$. The fit curves with and without 
the finite size effects are almost degenerate for 
$a m_{\rm ud}^{\rm  AWI}>0.003$, but we find  a sizable
difference at the physical point 
comparing the red open and solid symbols.
This feature is understood by Fig.~\ref{fig:ratio} where
we plot the magnitude of $R_X$ for $X=m_\pi,m_K,f_\pi,f_K$ 
with $L=2.8$fm as a function of $m_\pi$
( we note that $R_{m_{\rm PS}}>0$ and $R_{f_{\rm PS}}<0$); 
the finite size effects are less than 2\% for 
$m_{\rm PS}$ and $f_{\rm PS}$ at our simulation points.    
This is true even at the physical point 
except for $f_\pi$ which decreases by 4\%.

\begin{table}[t!]
\centering
\begin{tabular}{ccccc}\hline
$L_i(\mu=m_\eta)$ &PACS-CS  & PACS-CS with FSE      & exp. value\cite{Amoros:2001cp}  
& MILC\cite{Bernard:2006zp}
    \\
\hline
\hline
$L_4$        & 0.25(11)  &  0.23(12)      & 0.27 $\pm$ 0.8  & 0.1(2)(2)           \\
$L_5$        &  2.28(13) &  2.29(14)      & 2.28 $\pm$ 0.1  & 2.0(3)(2)           \\
$2L_6-L_4$ & 0.16(4)   &  0.16(4)          & 0 $\pm$ 1.0     & 0.5(1)(2)    \\ 
$2L_8-L_5$  & $-$0.59(5) & $-$0.60(5)         & 0.18 $\pm$ 0.5  & $-$0.1(1)(1)    \\
\hline
$\chi^2$/d.o.f. & 2.1(1.4) &  2.1(1.4) & &  \\
\hline
\end{tabular}
\caption{Results for the low energy constants together with the
phenomenological estimates and the MILC results.}
\label{tab:fit}
\end{table}

%
%


\subsection{Physical point and light hadron spectrum}
In order to determine the up-down and the strange quark masses and the
lattice cutoff we need three physical inputs.
We try the following two cases: 
$m_\pi, m_K, m_\Omega$ and $m_\pi, m_K, m_\phi$. 
The choice of $m_\Omega$ has theoretical and practical advantages:
the $\Omega$ baryon is stable in the strong interactions
and its mass, being composed of three strange quarks, 
is determined with good precision with small finite 
size effects. 
We also choose $m_\phi$ for comparison.
We employ the NLO ChPT formulae for the chiral extrapolations
of $m_\pi$, $m_K$, $f_\pi$ and $f_K$.  A simple linear formula
$m_{\rm had}=a+b\cdot m_{\rm ud}^{\rm AWI}+c\cdot m_{\rm s}^{\rm AWI}$ is
used for the other hadron masses, employing data in the same range   
$\kappa_{\rm ud}\ge 0.13754$ as for pseudoscalar mesons.
The results for the quark masses and the lattice cutoff are given by
\begin{eqnarray}
&&a^{-1}=2.256(81){\rm GeV}, \ \ \ m^{\overline{\rm MS}}_{ud}=2.37(11){\rm MeV},
\ \ \ m^{\overline{\rm MS}}_{s}=69.1(25){\rm MeV}, \ \ \  m_\Omega{\rm -input},\\
&&a^{-1}=2.248(76){\rm GeV}, \ \ \ m^{\overline{\rm MS}}_{ud}=2.38(11){\rm MeV},
\ \ \ m^{\overline{\rm MS}}_{s}=69.4(25){\rm MeV},  \ \ \  m_\phi {\rm -input}, \hspace{6mm}
\end{eqnarray}
where the errors are statistical. The two sets of results are
consistent within the error.  The quark masses
are smaller than the recent estimates in the literature.  We note, however, 
that we employed the perturbative renormalization factors to one-loop level 
which may contain  a sizable uncertainty.  A non-perutrbative calculation 
of the renormalization factor is in progress using the Schr\"odinger 
functional scheme.

Using the physical quark masses and the cutoff determined above, 
we obtain predictions for the pseudoscalar meson decay constants
at the physical point:
\begin{eqnarray}
&&f_\pi=144(6){\rm MeV}, \ \ \ f_K=175(6){\rm MeV}, \ \ \ f_K/f_\pi = 1.219(22), \ \ \  m_\Omega{\rm -input},\\
&&f_\pi=143(6){\rm MeV}, \ \ \ f_K=175(5){\rm MeV}, \ \ \ f_K/f_\pi = 1.219(21), \ \ \  m_\phi {\rm -input},\hspace{5mm}
\end{eqnarray} 
to be compared with the experimantal values 
$f_\pi=130.7$~MeV, $f_K=159.8$~MeV, $f_K/f_\pi = 1.223$.
A 10\% discrepancy in the magnitude of $f_\pi$ and $f_K$ 
might be due to use of one-loop perturbative $Z_A$ 
since the ratio shows a good agreement.  
A non-nonperturbative calculation of $Z_A$ and $Z_m=Z_A/Z_P$
is also in progress.

\begin{figure}[t!]
\vspace{3mm}
\begin{center}
\begin{tabular}{cc}
\includegraphics[width=70mm,angle=0]{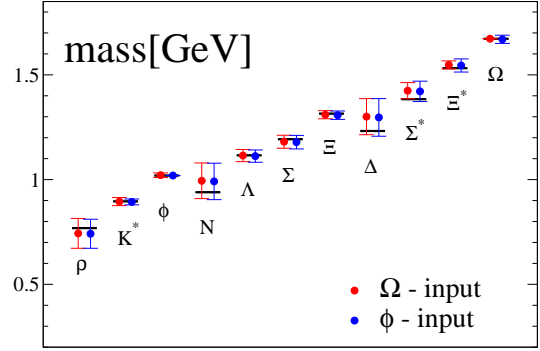}
\end{tabular}
\end{center}
\vspace{-.5cm}
\caption{Light hadron spectrum extrapolated to the physical point 
with  $\Omega$-input (red) and $\phi$-input (blue).
Horizontal bars denote the experimental values.}
\label{fig:spectrum}
\end{figure}


In Fig.~\ref{fig:spectrum} we compare the light hadron spectrum 
extrapolated to the physical point with the experiment.
The results for the $\Omega$-input and the $\phi$-input  
are consistent with each other, and both are in agreement 
with the experiment albeit errors are still not small 
for some of the hadrons. We find this to be encouraging.  
Further work is of course needed since 
cutoff errors of $O((a\Lambda_{\rm QCD})^2)$ are present 
in our results.

\section*{Acknowledgment}

This work is supported in part by Grants-in-Aid for Scientific
Research from the Ministry of Education, Culture, Sports, Science and
Technology (Nos. 13135204, 15540251, 17340066, 17540259,
18104005, 18540250, 18740139).

\end{document}